\documentclass{article}

\usepackage{arxiv}

\usepackage[utf8]{inputenc} 
\usepackage[T1]{fontenc}    
\usepackage{hyperref}       
\usepackage{url}            
\usepackage{booktabs}       
\usepackage{amsfonts}       
\usepackage{nicefrac}       
\usepackage{microtype}      
\usepackage{lipsum}		
\usepackage{graphicx}
\usepackage[numbers]{natbib} 
\usepackage{doi}

\usepackage{soul} 
\usepackage{xcolor}

\title{Evaluation of the Pronunciation of Tajweed Rules Based on DNN as a Step Towards Interactive Recitation Learning}

\date{} 					

\author{
  \href{https://orcid.org/0009-0006-3534-960X}{\includegraphics[scale=0.06]{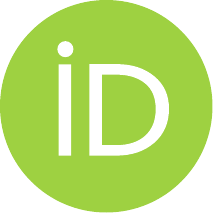}\hspace{1mm}Dim Shaiakhmetov} \\
  Department of Computer Science \\
  Ala-Too International University \\
  Bishkek, Kyrgyzstan \\
  \texttt{dim.shaiahmetov@alatoo.edu.kg} \\
  \And
  \href{https://orcid.org/0009-0001-9104-0918}{\includegraphics[scale=0.06]{orcid.pdf}\hspace{1mm}Gulnaz Gimaletdinova} \\
  Department of Applied Mathematics and Informatics \\
  Ala-Too International University \\
  Bishkek, Kyrgyzstan \\
  \texttt{gulnaz.gimaletdinova@alatoo.edu.kg} \\
  \And
  \href{https://orcid.org/0000-0003-0581-1913}{\hspace*{-5.5em}\includegraphics[scale=0.06]{orcid.pdf}\hspace{1mm}Selcuk Cankurt} \\
  \hspace*{-5.5em}Department of Computer Engineering \\
  \hspace*{-5.5em}Vistula University \\
  \hspace*{-5.5em}Warsaw, Poland \\
  \hspace*{-5.5em}Department of Engineering and Natural Sciences \\
  \hspace*{-5.5em}Suleyman Demirel University \\
  \hspace*{-5.5em}Almaty, Kazakhstan \\
  \hspace*{-5.5em}\texttt{s.cankurt@vistula.edu.pl} \\
  \And
  \href{https://orcid.org/0009-0002-9275-512X}{\hspace*{-2.5em}\includegraphics[scale=0.06]{orcid.pdf}\hspace{1mm}Kadyrmamat Momunov} \\
  \hspace*{-2.5em}Department of Computer Science \\
  \hspace*{-2.5em}Ala-Too International University \\
  \hspace*{-2.5em}Bishkek, Kyrgyzstan \\
  \hspace*{-2.5em}\texttt{kadyrmamatmomunov@gmail.com} \\
}

\begin{document}

\begin{titlepage}
    \centering
    \vspace*{4cm}
   
    {\large © 2025 IEEE.  Personal use of this material is permitted.  \\ 
    Permission from IEEE must be obtained for all other uses, in any current or future media, including reprinting/republishing this material for advertising or promotional purposes, creating new collective works, for resale or redistribution to servers or lists, or reuse of any copyrighted component of this work in other works.}

    \vspace{2em}

    {\large Related DOI: \href{https://doi.org/10.1109/CompSysTech65493.2025.11137272}{https://doi.org/10.1109/CompSysTech65493.2025.11137272}}

    \vfill
\end{titlepage}

\maketitle

\begin{abstract}
    Proper recitation of the Quran, adhering to the rules of Tajweed, is crucial for preventing mistakes during recitation and requires significant effort to master. Traditional methods of teaching these rules are limited by the availability of qualified instructors and time constraints. Automatic evaluation of recitation can address these challenges by providing prompt feedback and supporting independent practice. This study focuses on developing a deep learning model to classify three Tajweed rules—separate stretching (Al Mad), tight noon (Ghunnah), and hide (Ikhfaa)—using the publicly available QDAT dataset, which contains over 1,500 audio recordings. The input data consisted of audio recordings from this dataset, transformed into normalized mel-spectrograms. For classification, the EfficientNet-B0 architecture was used, enhanced with a Squeeze-and-Excitation block. The developed model achieved accuracy rates of 95.35\%, 99.34\%, and 97.01\% for the respective rules. An analysis of the learning curves confirmed the model's robustness and absence of overfitting. The proposed approach demonstrates high efficiency and paves the way for developing interactive educational systems for Tajweed study.
\end{abstract}



\section{Introduction}

The recitation of the Quran in accordance with Tajweed rules is an essential aspect of Muslim culture. Tajweed is a system of pronunciation that governs the articulation of Arabic sounds and their acoustic properties. These rules play a critical role in preserving the accurate recitation of the sacred text \cite{Tajweed_Mistakes}.

Traditionally, the process of teaching Quranic recitation is primarily carried out by qualified instructors with deep knowledge of Tajweed. This approach is effective for conveying the subtleties of pronunciation; however, it comes with several challenges. Access to experienced mentors is often limited. Furthermore, the evaluation of each student's pronunciation is done individually, in a one-on-one format, which is time-consuming. Lastly, the effectiveness of students' self-practice is reduced due to the lack of feedback during exercises. All of these factors create barriers for a wider audience interested in learning Quranic recitation.

The integration of artificial intelligence (AI) for the automated assessment of recitation can address many of these challenges. AI provides rapid pronunciation evaluation, enables the simultaneous instruction of multiple students, and enhances the effectiveness of self-practice through prompt feedback.

Moreover, the integration of AI with modern pedagogical approaches can further enhance these benefits. Research in the field of education shows that gamification and interactive learning applications significantly improve the quality of education by increasing student motivation and engagement \cite{criteria_of_assessing}. In this context, transforming the study and assessment of Tajweed rules into an interactive format, enhanced by AI capabilities \cite{QVoice}, can not only eliminate existing limitations but also elevate the educational process to a new level, making it more engaging and effective.

The objective of this study is to develop a deep learning model for automatic evaluation of pronunciation of three Tajweed rules — separate stretching (Al Mad), tight noon (Ghunnah), and hide {Ikhfaa} — using the publicly available QDAT dataset \cite{QDAT_dataset}.

This article is organized as follows. Section \ref{sec:2_lit_review} discusses existing research in the field of automatic evaluation of Tajweed rules, highlighting their approaches and limitations. Section \ref{sec:3_methodology} describes the process of developing the proposed model, including data preprocessing, architecture selection, and training parameters. Section \ref{sec:4_result} presents the achieved accuracy metrics and their analysis, as well as discusses the effectiveness of the approach. Finally, Section \ref{sec:5_conclusion} summarizes the main findings of the study.


\section{Literature Review}
\label{sec:2_lit_review}

Over the past decade, research on the automatic classification of Quranic recitation and Tajweed rules has been actively advancing, driven by progress in AI. These studies focus on analyzing audio data to support recitation training using various approaches.

Al-Marri et al. applied a combination of GMM-HMM and DNN for analyzing the pronunciation of non-native Arabic speakers during Quranic recitation \cite{16_Computer_aided}. Their DNN-GMM model for analyzing individual letters achieved an accuracy of 91.24\%, improving the baseline GMM-HMM by 0.78\%. Meftah et al. utilized an ANN with formant frequencies to recognize tight noon (Ghunnah), achieving an accuracy from 71.5\% to 85.4\% depending on the size of the dataset \cite{20_Identification}. This approach focused on a single Tajweed rule.

Deep learning methods have also gained widespread application. Al Harere et al. developed a system based on LSTM and MFCC for classifying three Tajweed rules — separate stretching, tight noon, and hide — on the publicly available QDAT dataset \cite{29_Mispronunciation_Detection}. Their model achieved accuracies of 96\%, 95\%, and 96\%, respectively, becoming a benchmark for our article. Alagrami et al. applied SVM with threshold scoring to analyze four rules (Edgham Meem, Ekhfaa Meem, Tafkheem Lam, Tarqeeq Lam), attaining an accuracy of 99\% \cite{30_Smartajweed}.

Recent studies also employ convolutional neural networks (CNNs) and transformers. Nazir et al. applied AlexNet with transfer learning to classify 28 Arabic phonemes, achieving an accuracy of 92.2\% \cite{28_Mispronunciation}. Sadik et al. used Vision Transformer (ViT) and transfer learning-based model (AlexNet) to analyze 2D spectrograms of children's voices, demonstrating average accuracy of 91.81\% \cite{33_2D_Spectrogram}. Asif et al. applied CNN for classifying short vowels, achieving 95.77\% accuracy after data augmentation \cite{9_An_approach}.

A review of existing research demonstrates a variety of approaches to classifying Quranic recitation audio data, including both traditional methods and deep learning. However, many studies rely on private datasets. In our work, we use the publicly available QDAT dataset. This aligns our research with study of \cite{29_Mispronunciation_Detection}, who also utilized this dataset. Consequently, their reported accuracies (96\%, 95\%, and 96\%) serve as a performance benchmark for our investigation.


\section{Methodology}
\label{sec:3_methodology}

\begin{figure}[ht]
  \centering
  \includegraphics[width=0.8\textwidth]{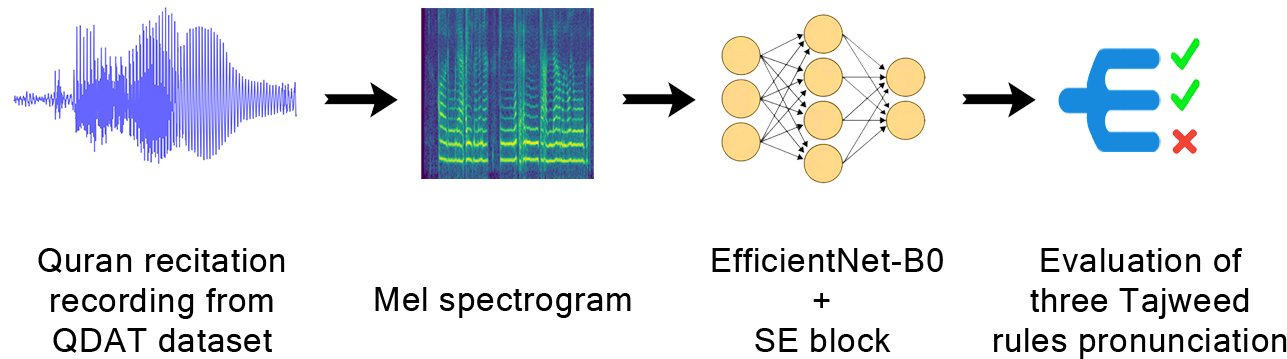}
  \caption{The proposed approach for automatic evaluation of the pronunciation of three Tajweed rules.}
  \label{fig:aproach_schema}
\end{figure}

This chapter is dedicated to the description of the approach for automatic classification of three Tajweed rules — separate stretching (Al Mad), tight noon (Ghunnah), and hide (Ikhfaa). \autoref{fig:aproach_schema} provides a brief overview of the developed method. The following subsections contain a detailed explanation of all stages of this process, including data preparation, model selection, and training.

\subsection{Dataset}

This study uses the QDAT dataset \cite{QDAT_dataset}, which was developed to evaluate methods for automatic detection of Quranic recitation rules. The dataset contains 1505 audio recordings of Quran recitations from Surah Al-Maidah verse 109, annotated according to three Tajweed rules: separate  stretching (Al Mad), tight noon (Ghunnah), and hide (Ikhfaa). Each rule is represented by a binary label: 0 for incorrect pronunciation and 1 for correct pronunciation, framing the task as a multiclass binary classification problem. The class distribution in the dataset is as follows: 43\% of the recordings have a label of 0 for separate stretching, 19\% for tight noon, and 47\% for hide. Consequently, the tight noon rule has an imbalance, with a predominance of correct pronunciation (81\%).

The objective of the study is to automatically evaluate pronunciation of three Tajweed rules in Quranic recitation. For each recording, the model must simultaneously predict three binary labels, corresponding to the correct application of the specified rules. The dataset is split into training and testing sets using a standard ratio of 80\% to 20\%. As our performance benchmark, we selected the results of Al Harere et al. \cite{29_Mispronunciation_Detection}, who achieved accuracies of 96\%, 95\%, and 96\% for the respective classes.

\subsection{Data Preprocessing}

To prepare the audio recordings from the QDAT dataset for training deep learning models, preprocessing was conducted to transform the audio data into a format suitable for analysis by CNN and transformers. The primary goal was to create uniform features that preserve the key characteristics of the Quranic recitation, without the use of data augmentation techniques.

During the data preparation process, the specific characteristics of the QDAT dataset were taken into account. While most recordings in the dataset were provided in WAV format, 20 of them were found to be in AC3 format. These were manually converted to WAV. Additionally, in the label table, we identified a missing label for file S22\_6. Based on the analysis of the remaining data from this speaker, it can be inferred that the missing value is likely 1, as all other values for this speaker are also 1.

Each audio recording was transformed into a mel-spectrogram—a two-dimensional representation that captures the frequency and temporal characteristics of the sound. The torchaudio library was used with a sampling rate of 11025 Hz. The transformation parameters included an Fast Fourier Transform (FFT) window size of 1024, a window hop of 256, and 224 mel filters, aligning with the 224×224×3 input size required by the EfficientNet-B0 architecture. The maximum frequency was limited to 4000 Hz, and the minimum frequency was set to 0. Initially, PitchShift augmentation was planned to increase data diversity; however, during experiments, this method was disabled as it did not improve classification results.

After obtaining the mel-spectrograms, they were normalized to stabilize the training process. The values of each image were logarithmically transformed with the addition of a small constant offset (\( 10^{-6} \)) to avoid computational errors. After that they were normalized by the mean and standard deviation. The final data format consisted of a tensor of size 224×224×3, where the three channels were formed by repeating the normalized spectrogram to ensure compatibility with the neural network architectures used in the project.

These preprocessing steps enabled the creation of a final set of audio recordings that were used for training the model.

\subsection{Initial Approaches}

At the initial stage of the study, several deep learning-based approaches were evaluated to address the task of Tajweed rule classification. These preliminary experiments helped identify limitations and set the direction for the subsequent development of the methodology.

The first approach involved an attempt to replicate the results described in \cite{29_Mispronunciation_Detection}, which utilized an LSTM network with MFCC features. Audio recordings from the QDAT dataset were converted into MFCCs with the parameters specified in \cite{29_Mispronunciation_Detection}: 40 Mel filters, a sampling rate of 11025 Hz, and 32 ms window duration. The LSTM architecture consisted of three layers, each with 256 neurons, as well as fully connected layers (16, 16, and 8 neurons) and final output layers for evaluating the pronunciation of the rules separate stretching, tight noon, and hide. However, during the experiments, the accuracy on the test set ranged between approximately 85\%-95\%, which was below the target benchmarks of 96\%, 95\%, and 96\%. This result was likely due to variations in training or preprocessing parameters, which were not comprehensively detailed in the paper \cite{29_Mispronunciation_Detection}.

Subsequently, CNNs were evaluated, which are better suited for processing two-dimensional data such as mel-spectrograms \cite{CNN_for_melspectr}. As our initial CNN architecture, we employed EfficientNet-B0 with all layers unfrozen to adapt model to the task. The training was conducted using the Adam optimizer and various loss weights to address class imbalance. The accuracy ranged between 90\% and 99\% for individual rules, but a consistent simultaneous surpassing of the target benchmarks was not observed.

Next, we experimented with MobileNetV2, EfficientNet-B1, ResNet-18, and VGG16 architectures, also with all layers unfrozen. EfficientNet-B1, MobileNetV2, and ResNet-18 showed results similar to EfficientNet-B0, with a maximum accuracy of around 94\%–96\% for individual rules, but without simultaneous improvement across all three rules. VGG16 was unable to achieve high performance, maintaining at an accuracy range of 60\%–80\% for all rules. These experiments demonstrated that standard CNNs require additional improvements to achieve higher and more stable results on this dataset.

Among all the tested architectures, EfficientNet-B0 demonstrated optimal performance in terms of training speed and accuracy. Therefore, it was selected as the baseline architecture for further refinement and experiments.

\subsection{Comparison of CNN Enhancement Methods}

To enhance the effectiveness of Tajweed rules classification,  attention mechanism and Squeeze-and-Excitation (SE) blocks were tested in addition to standard CNNs. These techniques help the model focus on the most significant features of the input data. While attention mechanisms dynamically identify key time-frequency regions in the mel-spectrogram, SE blocks adaptively rebalance channel strengths. In our case, these approaches were applied to mel-spectrograms to recognize specific characteristics of Quranic recitation, such as vowel prolongation or nasal tones, which are difficult to capture with a standard CNN \cite{CNN_without_attention}.

First, the Squeeze-and-Excitation (SE) block was applied to the EfficientNet-B0 architecture. Two configurations were tested: 1) inserting the SE block before Global Average Pooling, and 2) placing it after pooling. In both cases, the SE block enhanced the significant features of the data before passing them to the classifier. This led to an improvement in accuracy compared to the baseline model without SE block. 

Next, the Vision Transformer (ViT) approach was explored. It's a method that completely replaces convolutional layers with a transformer architecture. We employed a simplified version of ViT, where mel-spectrograms were split into patches and then processed using the Self-Attention mechanism. 

The SE block applied after pooling proved to be the most effective due to its simplicity of implementation and stable accuracy improvement on the small dataset, making it the optimal choice for training the model.

\subsection{Model Training}

The model training was conducted to classify three Tajweed rules: separate stretching, tight noon, and hide. After testing various models and attention mechanisms, the approach combining EfficientNet-B0 with the SE block after Global Average Pooling was selected. This choice allowed to achieve the best results on the QDAT dataset, which consists of 1505 audio recordings.

The EfficientNet-B0 architecture was employed with full layer unfrozen to adapt it to our classification task. After the convolutional layers, which produced features of size 1280×7×7, Global Average Pooling was applied, compressing the data into a vector of 1280 channels. Subsequently, the SE block processed this vector by calculating attention weights for each channel through a sequence of linear layers with a channel reduction (compression ratio of 16) and a Sigmoid activation function. These weights were multiplied by the original vector, enhancing the significant features. The final classifier included a Dropout layer with a probability of 0.7 and a fully connected layer with three outputs to evaluate the Tajweed rules.

Training was conducted using the Adam optimizer with a learning rate of 0.0001. To address the class imbalance for tight noon (19\% errors and 81\% correct recitations), loss weights of [1, 0.19, 0.95] were applied for the three rules, converted into positive weights as inverse values for the BCEWithLogitsLoss function. The training process spanned 40 epochs. As a result, the model achieved an accuracy of 95.35\% for separate  stretching, 99.34\% for tight noon, and 97.01\% for hide on the test set. The average accuracy was 97.23\%, surpassing the benchmark average accuracy of 95.33\% \cite{29_Mispronunciation_Detection}.


\section{Result and Discussion}
\label{sec:4_result}

This chapter presents the results of multiclass binary classification of three Tajweed rules and analyzes the effectiveness of the proposed approach.

We trained the model using audio recordings from the QDAT dataset, converted into normalized mel-spectrograms. The classification was performed using the EfficientNet-B0 architecture, enhanced with the Squeeze-and-Excitation block after Global Average Pooling.

The model achieved a pronunciation evaluation accuracy of 95.35\% for separate stretching (Al Mad), 99.34\% for  tight noon (Ghunnah), and 97.01\% for  hide (Ikhfaa) after 40 epochs of training. The average accuracy was 97.23\%, which is higher than the benchmark of 95.33\% set by \citet{29_Mispronunciation_Detection}, with results of 96\%, 95\%, and 96\% respectively.

\begin{figure}[htbp]
  \centering
  \includegraphics[width=0.8\textwidth]{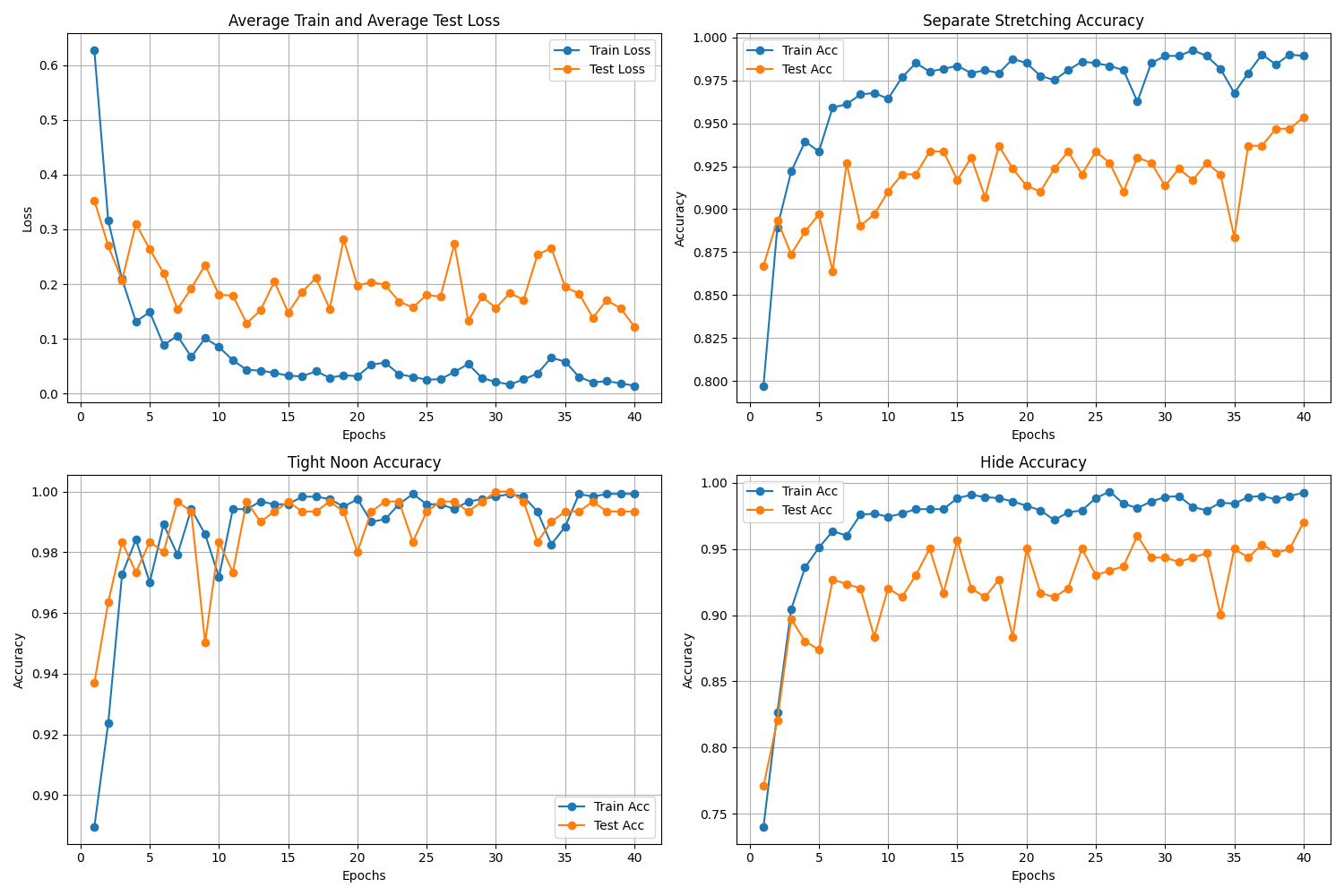}
  \caption{Learning curves}
  \label{fig:learning_curves}
\end{figure}

The analysis of the learning curves (see \autoref{fig:learning_curves}) confirmed the successful optimization. The final loss values for both the training (Train Loss) and test (Test Loss) sets were significantly lower than the initial values. Moreover, the final Test Loss value was the lowest observed throughout the entire training process. These results indicate effective model training.

No signs of overfitting were observed, as the test losses remained at a low level, and the accuracy on the test set showed a steady growth. The observed increases in loss at certain stages did not have a negative impact. These fluctuations were likely due to natural variations in the optimization process or the inherent complexity of the task, rather than model overfitting.

A significant improvement for tight noon (+4.34\%) was achieved through the application of loss weights that compensated for its class imbalance (19\% errors vs. 81\% correct), enabling the model to better recognize this class. The accuracy for hide increased by 1.01\%, confirming the effectiveness of the approach in identifying key features. However, for separate stretching, the result was lower by 0.65\%, which could be explained by the relatively small size of the dataset (1505 recordings), limiting the model's ability to generalize for all rules simultaneously.


\section{Conclusion}
\label{sec:5_conclusion}

This study focused on developing an effective approach for automatic evaluation of pronunciation of three Tajweed rules  — separate stretching (Al Mad), tight noon (Ghunnah), and hide (Ikhfaa). The research was conducted using the publicly available QDAT dataset. The task was framed as a multiclass binary classification problem, where three independent binary labels were predicted for each recording.

The data preprocessing process involved transforming the audio recordings into normalized mel-spectrograms, compatible with the tested neural network architectures. Employing mel-spectrograms as input data proved effective in preserving the temporal and frequency characteristics necessary for classifying Tajweed rules.

To identify the most suitable neural network architecture capable of effectively addressing the task, EfficientNet-B0, EfficientNet-B1, ResNet-18, VGG16, and MobileNetV2 were evaluated. The EfficientNet-B0 architecture was selected as the optimal choice in terms of training speed and accuracy. However, its standalone performance did not achieve the target benchmark. Consequently, additional testing was conducted with combinations of EfficientNet-B0 and attention mechanisms such as the SE block before pooling, the SE block after pooling, and Vision Transformer. The combination of EfficientNet-B0 with the SE block after pooling demonstrated the best performance. This configuration achieved an accuracies of 95.35\% for separate stretching (Al Mad), 99.34\% for tight noon (Ghunnah), and 97.01\% for hide (Ikhfaa). The average accuracy reached 97.23\%, surpassing the benchmark of 95.33\% established in the work of \citet{29_Mispronunciation_Detection}. The analysis of the learning curves, depicted in \autoref{fig:learning_curves}, confirmed successful optimization with no signs of overfitting.

Thus, the developed approach proved to be effective for automatic evaluation of pronunciation of Tajweed rules, demonstrating high accuracy and training stability. The obtained results open up prospects for further application of this configuration in tasks involving the processing of audio data with religious content, particularly in Quranic recitation analysis.

\bibliographystyle{unsrtnat}
\bibliography{references}  






\end{document}